\documentclass[11pt]{amsart}
\usepackage[english]{babel}
\usepackage{egothic}
\usepackage[T1]{fontenc}

\usepackage{amsmath}
\usepackage{amsthm}
\usepackage{amssymb}
\usepackage[all]{xy}
\usepackage{mathrsfs}
\usepackage{enumerate}
\usepackage{physics}
\usepackage[notcite, final, notref]{showkeys}
\usepackage{dsfont}
\usepackage{graphicx}
\usepackage{float}
\usepackage{xcolor}
\newtheorem{theorem}{Theorem}[section]
\newtheorem{problem}[theorem]{Problem}
\newtheorem{lemma}[theorem]{Lemma}
\newtheorem{proposition}[theorem]{Proposition}
\newtheorem{corollary}[theorem]{Corollary}
\newtheorem{definition}[theorem]{Definition}

\theoremstyle{definition}
\newtheorem{remark}[theorem]{Remark}

\newtheorem{example}[theorem]{Example}

\usepackage{xcolor}

\title[]{Superoscillations and the Klein-Gordon equation via the Fourier method}
\author{Kamal Diki}
\address{(KD)Clifford research group, Department of Electronics and Information Systems, Faculty of Engineering and Architecture, Ghent University, Krijgslaan 281, 9000 Ghent, Belgium}
\email{Kamal.Diki@UGent.be}
\author{Simon Verbruggen}
\address{(SV)Clifford research group, Department of Electronics and Information Systems, Faculty of Engineering and Architecture, Ghent University, Krijgslaan 281, 9000 Ghent, Belgium}
\email{Simon.Verbruggen@UGent.be}

\begin{document}
\maketitle
\begin{abstract}
We investigate the time-evolution problem associated with the Klein-Gordon equation, using superoscillations as initial data. Additionally, the Segal-Bargmann transform is used to derive integral representations of the resulting solutions.
\end{abstract}


\tableofcontents

\section{Introduction}
\setcounter{equation}{0}

The Schrödinger equation is obtained from the classical (non-relativistic) energy-momentum relation \cite{Schiff1968Quantum}. For a free particle, this relation is given by
\begin{equation*}
	E = \frac{\mathbf{p}^2}{2m},
\end{equation*}
where $m$ is the rest mass, and $\textbf{p}$ and $E$ denote the classical momentum and energy, respectively.
The following correspondence rules are then postulated, in which the momentum operator $\mathbf{p}_{\text{op}}$ and energy operator $E_{\text{op}}$, replace their classical counterparts
$$ E \to E_{\text{op}} = i \hbar \frac{\partial}{\partial t}, \quad 
\mathbf{p} \to \mathbf{p}_{\text{op}} = - i \hbar \nabla. $$
Substituting them into the classical energy-momentum relation, we get the Schrödinger equation:
\begin{equation*}
	i \hbar  \frac{\partial}{\partial t} \Psi = - \frac{\hbar^2}{2m} \nabla^2 \Psi.
\end{equation*}
Similarly, from the relativistic energy-momentum relation for a free, spinless particle
\begin{equation*}
	E^2 = m^2 c^4 + \mathbf{p}^2 c^2,
\end{equation*}
and applying the same correspondence rules, one obtains the Klein-Gordon equation
\begin{equation*}
	- \hbar^2 \frac{\partial^2}{\partial t^2} \Psi = m^2 c^4 \Psi - \hbar^2 c^2 \nabla^2 \Psi.
\end{equation*}
This is commonly written as
\begin{equation*}
	\left( \frac{1}{c^2} \frac{\partial^2}{\partial t^2} - \nabla^2 +  \frac{m^2 c^2}{\hbar^2} \right) \Psi = 0.
\end{equation*}
By converting to natural units 
\begin{equation*}
	c = 1 = \hbar,
\end{equation*}
the typical form of the Klein-Gordon equation in mathematical and theoretical physics literature is obtained:
\begin{equation}
	(\square + m^2) \Psi = 0,
\end{equation}
where $$\square:= \frac{\partial^2}{\partial t^2} - \Delta = \frac{\partial^2}{\partial t^2} - \nabla^2,$$ is known as the d'Alembert operator or wave operator.\\

The Klein-Gordon equation is thus the relativistic generalization of the Schrödinger equation for spinless particles.
However, unlike the Schrödinger equation, it does not admit a positive-definite quantity that can be interpreted as a probability density.
A natural incorporation of spin and a positive-definite probability density are only achieved for the Dirac equation.
Nevertheless, the Klein-Gordon equation remains fundamental in quantum field theory, where it arises as the field equation for scalar fields.
For an elementary treatment of the Schrödinger, Klein-Gordon and Dirac equation, we refer to Sections 6, 42 and 43 in \cite{Schiff1968Quantum} respectively.\\ 

On the other hand, Aharonov and his coauthors showed that measurements of a quantum mechanical system could be performed without disturbance of the system. The outcomes of these measurements, referred to in \cite{aav} as \textit{weak values}, extract information from the system without leading to a total wave function collapse.
The theory of superoscillations arose from this concept of a weak measurement of a quantum observable.
Mathematically, it refers to a linear combination of low-frequency Fourier components, whose spectrum is bounded, resulting in oscillations that can occur at frequencies beyond the spectral bound. The mathematics of superoscillations and its various applications in physics were developed and studied extensively; see \cite{ACSSTbook2017, ACSST2011JPA, BZACSSTRQHL2019}. 
 \\ 
 
The prototypical superoscillating function, that appears in the theory of weak values  is
\begin{equation}\label{FNEXP}
F_n(x,a)
=\sum_{j=0}^nC_j(n,a)e^{i(1-\frac{2j}{n})x},\ \ x\in \mathbb{R},
\end{equation}
where $a>1$ and the coefficients $C_j(n,a)$ are given by
\begin{equation}\label{Ckna}
C_j(n,a)=\binom{n}{j}\left(\frac{1+a}{2}\right)^{n-j}\left(\frac{1-a}{2}\right)^j.
\end{equation}
If we fix $x \in \mathbb{R}$  and we let $n$ tend to infinity, we obtain that
\begin{equation}
\label{limit}
\lim_{n \to \infty} F_n(x,a)=e^{iax},
\end{equation}
and the convergence is uniform on compact subsets of the real line. The term superoscillations refers to the fact that
 the Fourier components' frequencies $1-2j/n$ appearing in \eqref{FNEXP} are bounded by 1, while the limit function $e^{iax}$ has a frequency $a$ that can be arbitrarily larger than $1$. \\ \\ The Segal-Bargmann transform, originally introduced in \cite{Bargmann1961}, is of particular interest in quantum mechanics, as it unitarily maps the standard Hilbert space of square-integrable functions on the real line onto the Bargmann-Fock space, which consists of entire functions that are square-integrable with respect to the Gaussian measure on the complex plane. It turns out that the phenomenon of superoscillations can be connected to the theory of Fock spaces via the Segal-Bargmann transform; see \cite{ACDSS}. This connection has recently been extended to the case of the Short-time Fourier transform (STFT); see \cite{ACHA2024}. Moreover, new generating functions for superoscillatory coefficients, revealing interesting relations with special functions such as Bernstein and Hermite polynomials have been developed in \cite{CSSY2023}. Additionally, integral representations of superoscillations have been studied in \cite{BCSS2023}. Results on superoscillations using hypercomplex methods are discussed in \cite{CMPS2025}.  \\ \\
  Superoscillations have been studied extensively in non-relativistic quantum mechanics, as solutions to Schrödinger's equation \cite{ABCS2022, ACSSTbook2017, ACSST2011JPA, CGS2017}.  Recently, some papers have studied superoscillations for relativistic quantum mechanics, particularly under the Klein-Gordon equation. For example, the time-evolution and persistence of superoscillations under the three-dimensional Klein-Gordon equation were studied in \cite{CSST2020}.
In \cite{AG2024}, superoscillations were studied in the context of the Klein-Gordon equation and applied to quantum tunneling and scattering of particles, by constructing an extended Fock space.  \\ 

In this paper, we focus on the evolution of superoscillations under the one-dimensional Klein-Gordon equation. To address this challenge, we state the following two problems: 

\begin{problem}\label{NP1}
Let $a >1$ and $m\geq 0$. Study the time evolution problem associated with the Klein-Gordon equation given by:
\begin{equation}\label{ENP1}
\left(\displaystyle \frac{\partial^2}{\partial t^2} - \frac{d^2}{dx^2} + m^2\right) u(x,t)=L(x,t), 
\end{equation}
for every $(x,t)\in \mathbb{R}\times \mathbb{R}_+$. The initial conditions at time $t=0$ are given by 
\begin{equation}
u(x,0)=F_n(x,a), \quad \partial_t u(x,0)=\partial_{x}F_n(x,a).
\end{equation}
We now consider the following cases, corresponding to different source terms $L(x,t)$

\begin{enumerate}
 
\item The homogeneous problem corresponding to the source term
	\begin{equation*}\label{P1}
	L(x,t)=0.
	\end{equation*}
\item The Dirac excitation in space

 \begin{equation*}\label{P2}
	L(x,t)=\delta(x), 
	\end{equation*}
	where \( \delta(x) \) denotes the Dirac delta distribution centered at \( x = 0 \).
	
\item The Dirac excitation in space and time
	\begin{equation*}\label{P3}
	L(x,t)=\delta(x)\delta(t).
	\end{equation*}
\end{enumerate}
	
\end{problem}

\begin{problem}\label{NP2}

Let $a >1$ and $b>1$. Study the time evolution problem associated with the Klein-Gordon equation given by
\begin{equation*}
\left(\displaystyle \frac{\partial^2}{\partial t^2} - \frac{d^2}{dx^2} + m^2\right) u(x,t)=L(x,t), 
\end{equation*}
for every $(x,t)\in \mathbb{R}\times \mathbb{R}_+$. The initial conditions at time $t=0$ are given by 
\begin{equation}
u(x,0)=F_n(x,a), \quad \partial_t u(x,0)=F_n(x,b).
\end{equation}

We now consider the following cases, corresponding to different source terms $L(x,t)$

\begin{enumerate}
\item The homogeneous problem corresponding to the source term
	\begin{equation*}\label{P4}
	L(x,t)=0
	\end{equation*}
\item The Dirac excitation in space
 \begin{equation*}\label{P5}
	L(x,t)=\delta(x).
	\end{equation*}
\end{enumerate}

\end{problem}
\textit{Outline of the paper:}  In Section 2, we review fundamental definitions and properties of superoscillations and the Segal-Bargmann transform. 
Then, section 3 presents the main results of this paper. First, the time-evolution of the 1D Klein-Gordon equation is studied for different superoscillatory initial conditions as presented in Problems \ref{NP1} and \ref{NP2}. We also express the time evolution in terms of some infinite order differential operators.
Finally, the connection between the homogeneous solution obtained in Problem \ref{NP1} and the Segal-Bargmann transform is studied, which allows us to derive an integral representation for this solution, leading to novel integral representations for  the superoscillation sequence, and its derivative.  Section 4 is devoted to further discussions connecting our results with the covariance kernel function of Brownian motion.

\section{Methods}
\setcounter{equation}{0}

\subsection{Mathematics of superoscillations}
We will now review the basic poperties of superoscillations; for more details we refer to \cite{ACSSTbook2017}. Inspired by the prototypical example of superoscillations presented in \eqref{FNEXP}, it is possible to define the so-called {\em generalized Fourier sequences}. These are sequences of the form

\begin{equation}\label{basic_sequenceq}
f_n(x,a):= \sum_{j=0}^n Z_j(n,a)e^{ih_j(n)x},\ \ \ n\in \mathbb{N},\ \ \ x\in \mathbb{R},
\end{equation}
where $a\in\mathbb R$, $Z_j(n,a)$ and $h_j(n)$
are complex and real-valued functions, respectively.
The sequence \eqref{basic_sequenceq}
 is said to be {\em a superoscillating sequence} if
 $\displaystyle\sup_{j,n}|h_j(n)|\leq 1$ and
 there exists a compact subset of $\mathbb R$,
 which will be called {\em a superoscillation set},
 on which $f_n(x,a)$ converges uniformly to $e^{ig(a)x}$,
 where $g$ is a continuous real-valued function such that $|g(a)|>1$.

For the the prototypical example of superoscillations given by \eqref{FNEXP}, we observe how the value of the exponential function $x \to e^{iax}$ for large values of $x$ can be approximated as the limit of combinations of exponentials with frequencies $h_j(n)=1-\frac{2j}{n}$, as in the expression \eqref{FNEXP}. This phenomenon is a special case of a more general concept, known as the {\it supershift property}.

\begin{definition}[Supershift Property]\label{supershift}
 Let $\lambda\mapsto \varphi_\lambda(X)$ be a continuous complex-valued function
in the variable $\lambda \in \mathcal{I}$, where $\mathcal{I}\subseteq\mathbb{R}$ is an interval, and
$X\in \Omega $, where $\Omega$ is a domain.
We consider  $X\in \Omega $ as a parameter for the function
 $\lambda\mapsto \varphi_\lambda(X)$ where $\lambda \in \mathcal{I}$.
When  $[-1,1]$ is contained in $\mathcal{I}$ and $a\in\mathbb R$, we define the sequence
\begin{equation*}
\label{psisuprform}
\psi_{n,a}(X)=\sum_{j=0}^nC_j(n,a)\varphi_{1-\frac{2j}{n}}(X),
\end{equation*}
in which $\varphi_{\lambda}$ is computed just at the points $1-\frac{2j}{n}$ which belong to the interval $[-1,1]$ and the coefficients
$C_j(n,a)$ are defined for example as in \eqref{Ckna}, for $j=0,...,n$ and $n\in \mathbb{N}$.
If
$$
\lim_{n\to\infty}\psi_{n,a}(X)=\varphi_{a}(X)
$$
for $|a|>1$ arbitrary large (but belonging to $\mathcal{I}$), we say that the function
$\lambda\mapsto \varphi_{\lambda}(X)$,  for $X$ fixed, admits the supershift property.
\end{definition}

 \begin{remark}\label{anal-supershift}
Note that if the function $\varphi_\lambda (X)$ in Definition \ref{supershift} is analytic with respect to $\lambda$, then the function $\lambda\mapsto \varphi_\lambda(X)$ satisfies the supershift property. This can be viewed as a direct consequence of the result proved in \cite[Theorem 4.8]{ACJSSST2022} in the case of two variables. We refer to the recent works \cite{CSSY2025CAS, CSSY2025} for further discussions on the connection between the supershift property and analyticity.
 \end{remark}
 Finally, recall that the Fourier transform of a function $\phi \in L^1(\mathbb{R})$ is defined by
\begin{equation}
\widehat{\phi}(\omega)=\mathcal{F}\phi(\omega):= \int_\mathbb{R}e^{-i\omega t}\phi(t)dt, \quad \omega \in \mathbb{R}.
\end{equation}
This can be extended to tempered distributions $T\in \mathcal{S}'(\mathbb{R})$. The Fourier transform $\widehat{T}$ is then the tempered distribution given by
\begin{equation}
	\langle \widehat{T}, \varphi \rangle = \langle T, \widehat{\varphi} \rangle, \quad \forall \varphi \in \mathcal{S},
\end{equation}
where $\langle \cdot , \cdot \rangle$ denotes the duality pairing between $\mathcal{S}'(\mathbb{R})$ and $\mathcal{S}(\mathbb{R})$. Here $\mathcal{S}(\mathbb{R})$ denotes the Schwartz space of rapidly decreasing functions.

\begin{lemma}\label{Lemma: fourier of superoscillation}
 In the sense of distributions, we have
 \begin{equation}
  \widehat{F_n}(w,a)=2\pi \sum_{j=0}^{n}C_j(n,a)\delta_{1-\frac{2j}{n}}(w).
 \end{equation}
 \end{lemma}
 \begin{proof}
	In the space of the tempered distributions $\mathcal{S}'(\mathbb{R})$, the Fourier representation of the Dirac delta function is given by
	\begin{align*}
		\mathcal{F}(e^{ibx}) = 2 \pi \delta (w-b), \quad b\in \mathbb R.
	\end{align*}
	Applying this property to the terms of $F_n(x,a)$ we obtain
	\begin{align*}
		\widehat{F_n}(w,a)=2\pi \sum_{j=0}^{n}C_j(n,a)\delta_{1-\frac{2j}{n}}(w)
	\end{align*}
 \end{proof}

\subsection{The Segal-Bargmann transform} For a general introduction to the Fock space and Segal-Bargmann transform, we refer the reader to \cite{Folland,Hall2013,Ner,Zhu2012}. This space can be introduced via its geometric description as follows: 
\begin{definition}\label{Fockdef}
An entire function $f: \mathbb{C} \to \mathbb{C}$ belongs to the Fock space, denoted by $ \mathcal{F}(\mathbb{C})$, if
$$ \| f \|_{\mathcal{F}(\mathbb{C})}^2= \frac{1}{\pi} \int_{\mathbb{C}} | f(z)|^2 e^{-|z|^2} d \lambda(z) < \infty,$$
where $ d \lambda(z)=dx dy$ is the classical Lebesgue measure on $\mathbb{C}$ for $z=x+iy$.
\end{definition}
The space $\mathcal{F}(\mathbb{C})$ is equipped with the inner product

\begin{equation}
\langle f,g \rangle_{\mathcal{F}(\mathbb{C})}=\frac{1}{\pi}\int_\mathbb{C}\overline{g(z)}f(z)e^{-|z|^2} d \lambda(z),
\end{equation}
for all $f,g \in  \mathcal{F}(\mathbb{C})$. 

\begin{remark}\label{ProbaMeasure}
The measure $d\mu(z):=\frac{1}{\pi}e^{-|z|^2} d \lambda(z)$ is a probability measure in $\mathbb{C}$, as it satisfies

\begin{align*}
\displaystyle \mu(\mathbb{C})&=\frac{1}{\pi}\int_\mathbb{C} e^{-|z|^2} d \lambda(z)=1.\\
\end{align*}
\end{remark}
\begin{example}
The monomials $\displaystyle \lbrace z^k\text{ }| \text{ } k\in \mathbb{N}_0 \rbrace$ belong to the Fock space $\mathcal{F}(\mathbb{C})$ and form an orthogonal basis. Indeed, we have
$$\langle z^k,z^\ell\rangle_{\mathcal{F}(\mathbb{C})}= k!\delta_{k,\ell}, \quad k,\ell \in \mathbb{N}_0.$$
\end{example}
All the evaluation functionals on the Fock space are bounded. Therefore, by the Riesz representation theorem, one can prove that $\mathcal{F}(\mathbb{C})$ is a reproducing kernel Hilbert space. More precisely, we recall the following result:
\begin{theorem}
The Fock space is a reproducing kernel Hilbert space, with the reproducing kernel given by
\begin{equation}
K(z,w)=K_w(z):=e^{z \overline{w}}, \qquad \forall z,w \in \mathbb{C}.
\end{equation}
Moreover, the reproducing kernel property ensures that for every $f\in\mathcal{F}(\mathbb{C})$ and $w\in \mathbb{C} $, the value of $f(w)$ can be expressed as
\begin{equation}
f(w)=\displaystyle  \frac{1}{\pi}\int_{\mathbb{C}} \overline{K(z,w)} f(z) e^{-|z|^2}d \lambda(z).
\end{equation}
\end{theorem}
The normalized reproducing kernel associated with the Fock space is defined as
\begin{equation}
k(z,w)=k_{w}(z):=\frac{K(z,w)}{||K_w||_{\mathcal{F}(\mathbb{C})}}, \quad z\in\mathbb{C}.
\end{equation}

\begin{remark}
The explicit expression of the normalized reproducing kernel of the Fock space is given by
\begin{equation}
 k(z,w)=k_w(z)=e^{z\overline{w}-\frac{|w|^2}{2}}, \quad  z,w \in\mathbb{C}.
\end{equation}
\end{remark}

\begin{remark}
Note that the family of normalized Fock kernels $(k_w)_{w\in\mathbb{C}}$ is referred to as \textit{coherent states} in the terminology of quantum mechanics, as they form eigenstates of the annihilation operator. See \cite{Gaz} for further details and explanations. 
\end{remark}

 The \textit{Segal-Bargmann kernel} can be introduced and computed using the generating function of Hermite polynomials as follows:

\begin{definition}
The Segal-Bargmann kernel is defined by
$$\displaystyle   A(z,x)=A_z(x) := \sum_{k=0}^{\infty}\psi_k(x)\frac{\overline{z}^k}{\sqrt{k!}}=\pi^{-1/4} e^{-\frac{1}{2}(x^2+\overline{z}^2)+\sqrt{2}\overline{z}x},$$
for every $z\in \mathbb{C}$ and $x\in\mathbb{R}$. Here $(\psi_k)_{k\in \mathbb{N}_0}$ denote the normalized Hermite functions.
\end{definition}
\begin{remark}
The reproducing kernel of the Fock space can be factorized in terms of the inner product in $L^2(\mathbb{R})$ using the Segal-Bargmann kernel, leading to

$$\langle A_w, A_z \rangle_{L^2(\mathbb{R})}=e^{z\overline{w}}=K(z,w), \quad \forall z,w\in \mathbb{C}.$$
\end{remark}

The Segal-Bargmann transform, introduced in \cite{Bargmann1961}, is defined as follows:

\begin{definition}
The Segal-Bargmann transform of a function $\psi\in L^2(\mathbb{R})$ is given by

\begin{align*}
\displaystyle (\mathcal{B}\psi)(z)&:=\langle \psi, A_{z} \rangle_{L^2(\mathbb{R})} \\
&=\pi^{-1/4} \int_{\mathbb{R}} e^{-\frac{1}{2}(x^2+z^2)+\sqrt{2}zx}\psi(x)dx,
\end{align*}
where $z\in \mathbb{C}$ and $A_z$ is the Segal-Bargmann kernel.
\end{definition}

The Segal-Bargmann transform establishes a connection between the Schrödinger representation and the Bargmann-Fock representation through the fundamental result obtained in \cite{Bargmann1961}. Furthermore, $\mathcal{B}$ transforms the normalized Hermite functions $(\psi_k)_{k\in\mathbb{N}_0}$ onto an orthonormal basis of the Fock space $\mathcal{F}(\mathbb{C})$, given by
\begin{equation}
(\mathcal{B}\psi_k)(z)=\frac{z^k}{\sqrt{k!}}:=e_k(z), \quad k\in \mathbb N_0.
\end{equation}

\section{Results}
\setcounter{equation}{0}

\subsection{The Klein-Gordon equation and superoscillations}
We start by considering the homogeneous case in Problem \ref{NP1}, we will work in $\mathcal{S}'(\mathbb{R})$ and apply the Fourier transform with respect to $x$ to equation \eqref{ENP1} with $L(x,t)=0$. Using the properties of the Fourier transform, we obtain the following ordinary differential equation (O.D.E):
 
 \begin{equation}\label{E2}
 \displaystyle \frac{\partial^2}{\partial t^2}\widehat{u}(w,t)=-(m^2+w^2)\widehat{u}(w,t).
 \end{equation}
 \begin{lemma}\label{Lemma general solution 1D homogenous klein gordon}
 The general solution of the equation \eqref{E2} is expressed as
 $$\widehat{u}(w,t)=A(w)\cos(\sqrt{m^2+w^2}t)+B(w)\sin(\sqrt{m^2+w^2}t).$$

 \end{lemma}
 \begin{proof}
 This is a homogeneous O.D.E. which can be solved using its characteristic equation
 \begin{align*}
 r^2 + (m^2 + w^2)=0
 \end{align*}
 Solving for $r$, we find $r=\pm i\sqrt{m^2 + w^2}$. Taking a linear combination of the corresponding solutions $e^{rx}$, we obtain the general solution
 \begin{align*}
	\widehat{u}(w,t)=A(w)\cos(\sqrt{m^2+w^2}t)+B(w)\sin(\sqrt{m^2+w^2}t)
 \end{align*}
 \end{proof}
 \begin{remark}\label{Remark wm expression}
 Sometimes, we use the notation $w_m=\sqrt{m^2+w^2}$ for  every $m, w\in\mathbb{R}$.
 \end{remark}
 \begin{lemma}\label{Lemma coefficients problem 3.1}
 The constants $A(w)$ and $B(w)$ in Lemma \ref{Lemma general solution 1D homogenous klein gordon} are determined using the initial conditions and are given by 
 
 $$A(w)=\widehat{F_n}(w,a)=2 \pi \sum_{j=0}^{n} C_j(n,a) \delta_{1- \frac{2j}{n} } , $$
 and  $$B(w)=i\frac{w}{\sqrt{m^2+w^2}}\widehat{F_n}(w,a)=2 \pi \frac{i w}{w_m} \sum_{j=0}^{n} C_j(n,a) \delta_{1-\frac{2j}{n}} ,$$
 with $w_m=\sqrt{m^2+w^2}$ for every  $w\in \mathbb R$.
 \end{lemma}
 \begin{proof}
	After applying the Fourier transform to the initial condition $u(x,0)=F_n(x,a)$, $A(w)$ can be determined as
	
		$$\displaystyle\hat{u}(w, 0) = \hat{F_n}(w,a), $$
		leading to 
		
		$$\displaystyle  A(w) =\hat{F_n}(w,a).$$

	From Lemma \ref{Lemma: fourier of superoscillation}, it follows that
	
	\begin{align*}
	A(w)&=2 \pi \sum_{j=0}^{n} C_j(n,a) \delta \left(w- \left(1- \frac{2j}{n} \right) \right)\\
	&= 2 \pi \sum_{j=0}^{n} C_j(n,a) \delta_{1- \frac{2j}{n} }.\\
	\end{align*}

	Simlarly, $B(w)$ can be determined from $\partial_t u(x,0)=\partial_{x}F_n(x,a)$. First we compute $\frac{\partial}{\partial t}\hat{u}(w, 0)$:
	\begin{align*}
		\frac{\partial}{\partial t}\hat{u}(w, 0) = - \left. A(w) w_m \sin(w_m t)+B(w) w_m \cos(w_m t) \right|_{t=0} = B(w) w_m.
	\end{align*}
	Applying the Fourier transform to the initial condition then yields:

	$$\frac{\partial}{\partial t}\hat{u}(w, 0) = \mathcal{F}[\partial_{x}F_n(x,a)],$$
	leading to 
	
	$$B(w) w_m =  i w \hat{F_n}(w,a).$$
	
Finally, we obtain
\begin{align*}
B(w) &=  2 \pi \frac{i w}{w_m} \sum_{j=0}^{n} C_j(n,a) \delta \left(w- \left(1- \frac{2j}{n} \right) \right) \\
&=  2 \pi \frac{i w}{w_m} \sum_{j=0}^{n} C_j(n,a) \delta _{1- \frac{2j}{n}},\\
\end{align*}
	
where $w_m=\sqrt{m^2+w^2}$ for every  $w\in \mathbb R$.
 \end{proof}
 
\begin{theorem}\label{Theorem: time evolution P1}
The time evolution of the superoscillating sequence $F_n(x,a)$ for the 1D-Klein-Gordon equation considered in the first case of Problem \ref{NP1} is given by 
\begin{equation}\label{Equation: time evolution problem 3.1}
\begin{aligned}
u_n(x,t) = \sum_{j=0}^{n} C_j(n,a) e^{ix(1-\frac{2j}{n})} \Bigg[ &\cos\left(\sqrt{m^2+\left(1-\frac{2j}{n}\right)^2} \, t\right) \\
&+ i \frac{1-\frac{2j}{n}}{\sqrt{m^2+\left(1-\frac{2j}{n}\right)^2}} \sin\left(\sqrt{m^2+\left(1-\frac{2j}{n}\right)^2} \, t\right) \Bigg].
\end{aligned}
\end{equation}

If $u(x,t)$ denotes the limit case when $n$ goes to $\infty$, we have

\begin{equation}\label{Equation: limitcase time evolution problem 3.1}
u(x,t)= e^{iax}\left[\cos(\sqrt{m^2+a^2}t)+i\frac{a}{\sqrt{m^2+a^2}}\sin(\sqrt{m^2+a^2}t)\right].
\end{equation}

\end{theorem}

\begin{proof}
	Substituting the coefficients $A(w)$ and $B(w)$ from Lemma \ref{Lemma coefficients problem 3.1} in the general solution from Lemma \ref{Lemma general solution 1D homogenous klein gordon} gives
	\begin{align*}
		\hat{u}(w,t) = 2 \pi \sum_{j=0}^{n} C_j(n,a) \delta \left(w- \left(1- \frac{2j}{n} \right) \right) \left[ \text{cos}(w_m t) + \frac{iw}{w_m} \text{sin}(w_m t) \right]
	\end{align*}
	Applying the inverse Fourier transform and inserting the expression for $w_m$ (Remark \ref{Remark wm expression}), we have
	\begin{align*}
		u(x,t) &= \sum_{j=0}^{n} C_j(n,a) \left[ \int_{\mathbb{R}} \delta \left(w- \left(1- \frac{2j}{n} \right) \right) \text{cos}(\sqrt{m^2 + w^2} t)e^{i w x} dw \right. \\
    	& \left. + \int_{\mathbb{R}} \delta \left(w- \left(1- \frac{2j}{n} \right) \right) \frac{iw}{w_m} \text{sin}(\sqrt{m^2 + w^2} t)e^{i w x} dw \right]
	\end{align*}
	We arrive at the expression for the time evolution
	\begin{align*}
		u_n(x,t) = \sum_{j=0}^{n} C_j(n,a) e^{ix(1-\frac{2j}{n})} \Bigg[ &\cos\left(\sqrt{m^2+\left(1-\frac{2j}{n}\right)^2} \, t\right) \\
		&+ i \frac{1-\frac{2j}{n}}{\sqrt{m^2+\left(1-\frac{2j}{n}\right)^2}} \sin\left(\sqrt{m^2+\left(1-\frac{2j}{n}\right)^2} \, t\right) \Bigg].
	\end{align*}
The limit case can be justified using the supershift property as discussed in Remark \ref{anal-supershift}. This follows from the analyticity of the function $u(x,t)$ with respect to the variable $a$.
\end{proof}

\begin{remark}
A plot of the superoscillating sequence $F_n(x,a)$ can be found in \cite{AG2024}. It was observed there that $F_n(x,a)$ could be approximated by its limit function $F(x,a)=e^{iax}$ near the origin, specifically within the range of $|x| < \sqrt{n}$.
The convergence to the limit function $F(x,a)$, as a function of $n$, is slower for higher values of $a$.
Several examples of the solution of Theorem \ref{Theorem: time evolution P1} are shown in Figure 1. In this Figure, we set $m=3$ and $n=10$ and plot examples for different values of $a$ as a function of $x$, for a fixed $t=0$. 
When looking at the zoomed-in subplot over the range $x=-2 \to 2$, we observe something similar to what was reported in \cite{AG2024}: near the origin, the superoscillating functions oscillate faster than their maximal Fourier components, which are bounded by $1$, and thus oscillate with the maximum frequency of following function
\begin{equation*}
    u(x,t) |_{a=1}= e^{ix}.
\end{equation*}
In the vicinity of $x=0$, the functions $u_n(x,t)$ may be approximated by their limit function $u(x,t)$. The convergence again visually deteriorates for higher $a$ values.

\begin{figure}[H]
    \centering
    \includegraphics[width=0.8\linewidth]{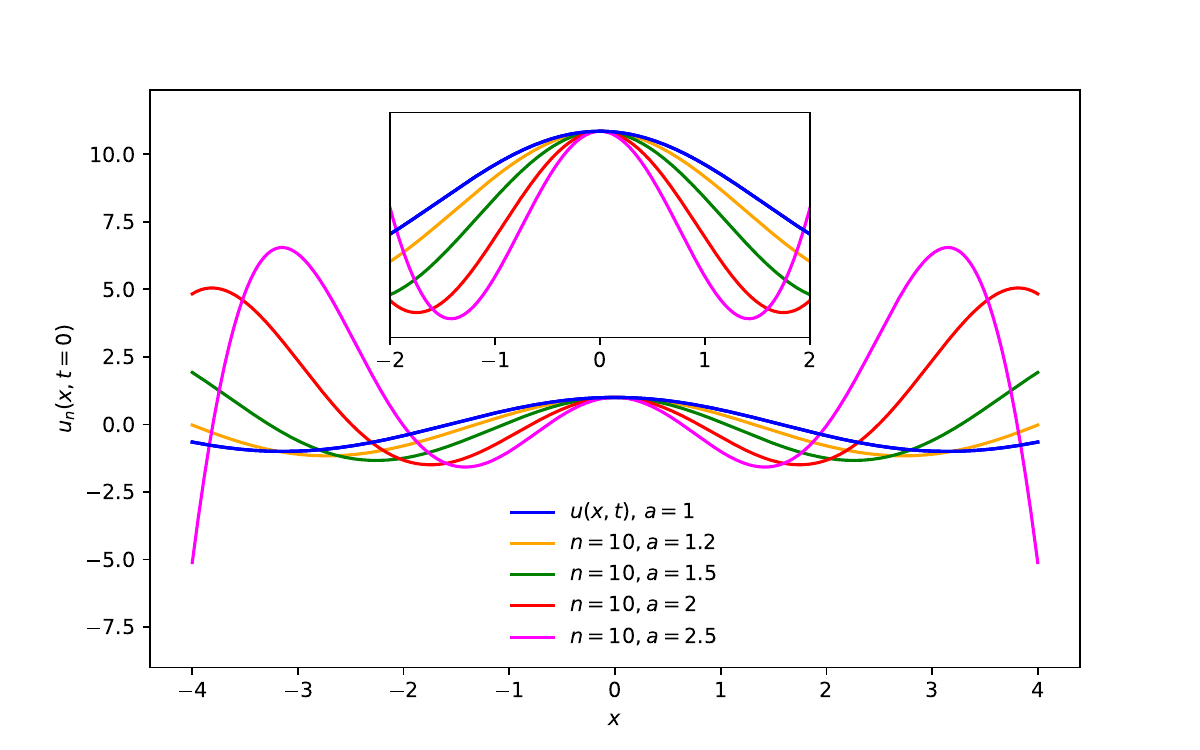}
    \label{fig:theorem 3.4}
	\caption{Plot of solution $u_n(x,t)$ of Theorem \ref{Theorem: time evolution P1} for $m=3$ and $n=10$. The solution is plotted for different values of $a$ as a function of $x$, for a fixed $t=0$.}
\end{figure}
\end{remark}

\begin{figure}[H]
    \centering
    \includegraphics[width=0.8\linewidth]{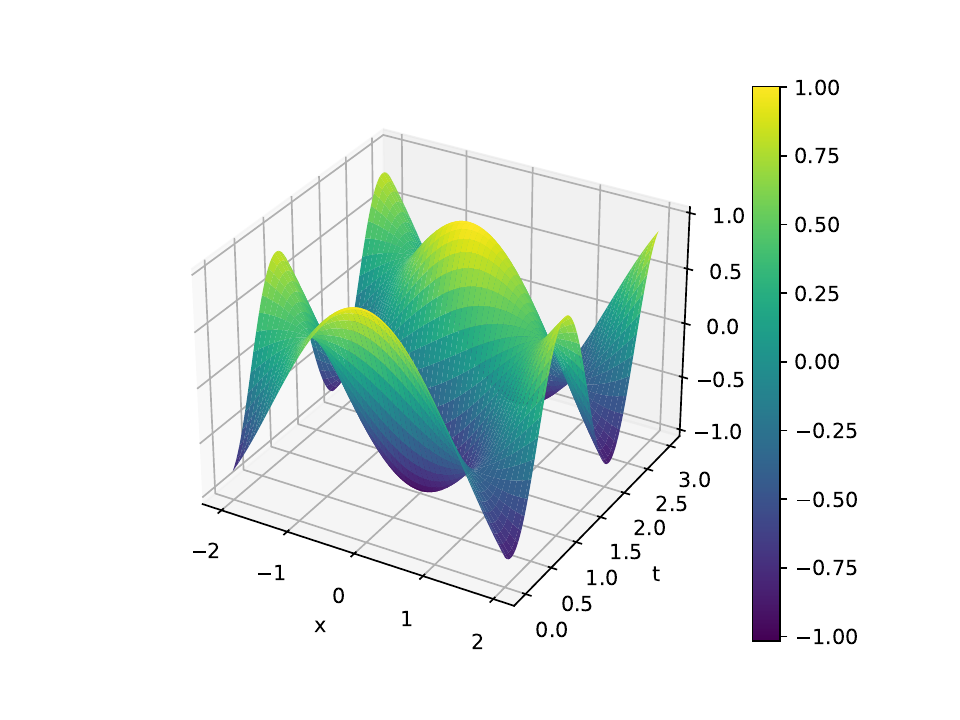}
    \caption{Two-dimensional plot of the evolution in time $u_n(x,t)$, for the example of $n=10, a=1.5, m=3$.}
    \label{fig: 2D theorem 3.4}
\end{figure}

An example of a two-dimensional plot of the evolution in time of a certain superoscillating sequence is displayed in Figure \ref{fig: 2D theorem 3.4}.

\begin{corollary}

If $m=0$,  the Klein-Gordon equation reduces to

$$\displaystyle \frac{\partial^2}{\partial t^2}u(x,t)=\frac{d^2}{dx^2} u(x,t).$$
In this case, we have 

\begin{equation}
u_n(x,t)=F_n(x+t,a):=\exp(t\frac{\partial}{\partial x})F_n(x,a),
\end{equation}
 and 
\begin{equation}
 u(x,t)=e^{ia(x+t)}=F(x+t,a)=\exp(t\frac{\partial}{\partial x})F(x,a).
\end{equation} 

\end{corollary}
\begin{proof}
	For $m=0$, the time evolution from Theorem \ref{Theorem: time evolution P1} reduces to:
	\begin{align*}
		u_n(x,t) &= \sum_{j=0}^{n} C_j(n,a)  e^{ix(1-\frac{2j}{n})} \Bigg[ \cos\left(\left(1-\frac{2j}{n}\right) \, t\right)
		+ i  \sin\left(\left(1-\frac{2j}{n}\right)\, t \right) \Bigg] \\
		 &= \sum_{j=0}^{n} C_j(n,a) e^{ix(1-\frac{2j}{n})} e^{it(1-\frac{2j}{n})} \\
		& = F_n(x+t,a).\\
	\end{align*}
	Using the expansion
	\begin{align*}
		e^{it(1-\frac{2j}{n})} = \sum_{k=0}^{\infty} \frac{t^k}{k!} \left[ i \left(1-\frac{2j}{n}\right) \right]^{k}
	\end{align*}
	The function $u_n(x,t)$ can be rewritten as
	\begin{align*}
		u_n(x,t) &= \sum_{k=0}^{\infty} \frac{t^k}{k!} \frac{\partial^k}{\partial x^k} \sum_{j=0}^{n} C_j(n,a) e^{ix(1-\frac{2j}{n})} \\
		         &= \exp(t\frac{\partial}{\partial x})F_n(x,a)
	\end{align*}
	Using the limit for $F_n(x,a)$ (eq. \ref{limit}), we now obtain

		$$u(x,t)= \lim_{n \to \infty} u_n(x,t) = \lim_{n \to \infty} F_n(x + t,a)=e^{ia(x+t)}. $$
		
	Therefore, we get 
	$$u(x,t)= \sum_{k=0}^{\infty} \frac{t^k}{k!} \frac{\partial^k}{\partial x^k} e^{iax} = \exp(t\frac{\partial}{\partial x}) F(x,a).$$	
\end{proof}

\begin{theorem}
The time evolution $u_n(x,t)$ associated with the first case of Problem \ref{NP1} can be expressed in terms of two infinite order differential operators as follows
\begin{equation}
	\displaystyle u_n(x,t) =  \sum_{j=0}^{n} C_j(n,a) \left( \mathcal{U}_1(t,D_x) + D_{x} \mathcal{U}_2(t,D_x) \right) e^{ix(1-\frac{2j}{n})}
\end{equation}
where the infinite order differential operators $\mathcal{U}_1$, and $\mathcal{U}_2$ are defined by
\begin{align*}
	\mathcal{U}_1(t,D_x) &= \cos\left( \sqrt{ m^2+\left(\frac{D_{x}}{i} \right)^2} t \right):= \sum_{n=0}^{\infty} \frac{(-1)^n}{(2n)!} \left[ m^2+\left(\frac{D_{x}}{i} \right)^2  \right]^{n} t^{2n}, 		\\
	\text{and} \\
	\mathcal{U}_2(t,D_x) &= \sin\left( \sqrt{ m^2+\left(\frac{D_{x}}{i} \right)^2} t \right):= \sum_{n=0}^{\infty} \frac{(-1)^n}{(2n+1)!} \left[ m^2+\left(\frac{D_{x}}{i} \right)^2  \right]^{n} t^{2n+1}. 	\\
\end{align*}
Here we use the notation $D_{x} := \frac{d}{d x}$.
\begin{proof}
	Applying the subsitution $y = \sqrt{m^2+\left(1 - \frac{2j}{n} \right)^2}t$ in the equation \ref{Equation: time evolution problem 3.1} yields
	\begin{equation*}
		u_n(x,t) = \sum_{j=0}^{n} C_j(n,a) e^{ix(1-\frac{2j}{n})} \left[ \cos (y) + i \left(1-\frac{2j}{n}\right) \frac{t}{y} \sin(y) \right]
	\end{equation*}
	We recall the Taylor series for sine and cosine
$$\sin(y) = \sum_{n=0}^{\infty} \frac{(-1)^n}{(2n+1)!} y^{2n+1}, \text{ and }\cos(y) = \sum_{n=0}^{\infty} \frac{(-1)^n}{(2n)!} y^{2n}. $$

	Substituting these expressions, we get
	\begin{equation*}
		u_n(x,t) = \sum_{j=0}^{n} C_j(n,a) \left[ \sum_{n=0}^{\infty} \frac{(-1)^n}{(2n)!} y^{2n} + i \left(1-\frac{2j}{n}\right) \frac{t}{y} \sum_{n=0}^{\infty} \frac{(-1)^n}{(2n+1)!} y^{2n+1} \right] e^{ix(1-\frac{2j}{n})}
	\end{equation*}
	Using the taylor series expansion of the exponential and substituting $y = \sqrt{m^2+\left(1 - \frac{2j}{n} \right)^2}t$, we obtain
	\begin{align*}
		u_n(x,t) &= \sum_{j=0}^{n} C_j(n,a) \left[ \sum_{n=0}^{\infty} \frac{(-1)^n}{(2n)!} \left( m^2+\left(1-\frac{2j}{n}\right)^2 \right)^{n} t^{2n} \right. \\
        & \left. + i \left(1-\frac{2j}{n}\right) \sum_{n=0}^{\infty} \frac{(-1)^n}{(2n+1)!}  \left( m^2+\left(1-\frac{2j}{n}\right)^2 \right)^{n} t^{2n+1} \right] e^{ix(1-\frac{2j}{n})}
	\end{align*}
	We now remark that the exponential function $e^{ix(1-\frac{2j}{n})}$ is an eigenfunction of $D_{x}$ with eigenvalue $i\left(1-\frac{2j}{n}\right)$
	\begin{equation*}
		D_{x} e^{ix(1-\frac{2j}{n})} = i\left(1-\frac{2j}{n}\right) e^{ix(1-\frac{2j}{n})}.
	\end{equation*}
	To apply this to the differential operators $\mathcal{U}_1$ and $\mathcal{U}_2$, we consider the binomial expansion of $\left(m^2 + \left(1-\frac{2j}{n}\right)^2\right)^n$
	\begin{equation*}
		\left(m^2+\left(1-\frac{2j}{n}\right)^2\right)^n = \sum_{k=0}^{n} {n \choose k}  m^{2k} \left(1-\frac{2j}{n}\right)^{2n-2k} 
	\end{equation*}
	Using again that $e^{ix(1-\frac{2j}{n})}$ is an eigenfunction of $D_x$, we get
	\begin{align*}
		\left[ \left(m^2+\left(1-\frac{2j}{n}\right)^2\right)^n \right] e^{ix(1-\frac{2j}{n})} &= \left[ \sum_{k=0}^{n} {n \choose k} m^{2k} \left(1-\frac{2j}{n}\right)^{2n-2k}  \right] e^{ix(1-\frac{2j}{n})} \\
		& = \left[ \sum_{k=0}^{n} {n \choose k} m^{2k} \left( \frac{D_{x}}{i} \right)^{2n-2k}  \right] e^{ix(1-\frac{2j}{n})} \\
		& = \left[ \left( m^2 + \left(\frac{D_{x}}{i} \right)^2 \right)^{n} \right] e^{ix(1-\frac{2j}{n})}.
	\end{align*}
	This implies that
	\begin{align*}
		\displaystyle  \left[ \sum_{n=0}^{\infty} \frac{(-1)^n}{(2n)!} \left( m^2 + \left(1-\frac{2j}{n}\right)^2  \right)^{n} t^{2n} \right] e^{ix(1-\frac{2j}{n})} &= \left[ \sum_{n=0}^{\infty} \frac{(-1)^n}{(2n)!} \left( m^2+\left(\frac{D_{x}}{i} \right)^2  \right)^{n} t^{2n} \right] e^{ix(1-\frac{2j}{n})} \\ 
		 & = \mathcal{U}_1(t,D_x)  e^{ix(1-\frac{2j}{n})},\\
		\end{align*}
	and 	
	
	\begin{align*}
	\displaystyle	 \left[ \sum_{n=0}^{\infty} \frac{(-1)^n}{(2n+1)!}  \left( m^2+\left(1-\frac{2j}{n}\right)^2 \right)^{n} t^{2n+1} \right] e^{ix(1-\frac{2j}{n})} &= \left[ \sum_{n=0}^{\infty} \frac{(-1)^n}{(2n+1)!} \left( m^2+\left(\frac{D_{x}}{i} \right)^2  \right)^{n} t^{2n+1} \right] e^{ix(1-\frac{2j}{n})}\\ 
	& = \mathcal{U}_2(t,D_x)  e^{ix(1-\frac{2j}{n})}.\\
		  \end{align*}

	Using these properties, we arrive at
	\begin{align*}
		u_n(x,t)  &=  \sum_{j=0}^{n} C_j(n,a)  \left[  \sum_{n=0}^{\infty} \frac{(-1)^n}{(2n)!} \left( m^2+\left(\frac{D_{x}}{i} \right)^2 \right)^{n} t^{2n} \right. \\
        & \left. + i \left(\frac{D_{x}}{i} \right) \sum_{n=0}^{\infty} \frac{(-1)^n}{(2n+1)!} \left( m^2+\left(\frac{D_{x}}{i} \right)^2 \right)^{n} t^{2n+1} \right] e^{ix(1-\frac{2j}{n})} \\
		& = \sum_{j=0}^{n} C_j(n,a) \left[ \mathcal{U}_1(t,D_x) + D_{x} \mathcal{U}_2(t,D_x) \right] e^{ix(1-\frac{2j}{n})}
	\end{align*}
\end{proof}

\end{theorem}

\begin{theorem}\label{P2-R1}
The time evolution of the superoscillating sequence $F_n(x,a)$ for the 1D-Klein-Gordon equation considered in the second case of Problem \ref{NP1} is given by 
\begin{equation}
\begin{aligned}
v_n(x,t) = \sum_{j=0}^{n} C_j(n,a) e^{ix(1-\frac{2j}{n})} \Bigg[ &\cos\left(\sqrt{m^2+\left(1-\frac{2j}{n}\right)^2} \, t\right) \\
&+ i \frac{1-\frac{2j}{n}}{\sqrt{m^2+\left(1-\frac{2j}{n}\right)^2}} \sin\left(\sqrt{m^2+\left(1-\frac{2j}{n}\right)^2} \, t\right) \Bigg] \\
&+\frac{1}{2\pi} \int_{\mathbb R} e^{ix \omega }\frac{\left(1-\cos(\sqrt{m^2+\omega^2}t)\right)}{m^2+\omega^2}d\omega.
\end{aligned}
\end{equation}

If $v(x,t)$ denotes the limit case when $n$ goes to $\infty$, we obtain 

\begin{equation}
\begin{aligned}
v(x,t)= e^{iax}&\left(\cos(\sqrt{m^2+a^2}t)+i\frac{a}{\sqrt{m^2+a^2}}\sin(\sqrt{m^2+a^2}t)\right) \\ &+\frac{1}{2\pi} \int_{\mathbb R} e^{ix \omega }\frac{\left(1-\cos(\sqrt{m^2+\omega^2}t)\right)}{m^2+\omega^2}d\omega.
\end{aligned}
\end{equation}

\end{theorem}
\begin{proof}
	Following the same approach as in Problem \ref{P1}, we begin by applying the Fourier transform to the equation (\ref{E2}), which yields the following O.D.E.: 
	\begin{align*}
		\displaystyle \frac{\partial^2}{\partial t^2}\widehat{u}(w,t)+(m^2+w^2)\widehat{u}(w,t) = 1.
	\end{align*}
	To solve this initial value problem (IVP), we need the solution to the homogeneous IVP (we denote this solution as $u_{n,h}$) which is given by Theorem \ref{Theorem: time evolution P1}:
	\begin{align*}
		u_{n,h}(x,t) = \sum_{j=0}^{n} C_j(n,a) e^{ix(1-\frac{2j}{n})} \Bigg[ &\cos\left(\sqrt{m^2+\left(1-\frac{2j}{n}\right)^2} \, t\right) \\
&+ i \frac{1-\frac{2j}{n}}{\sqrt{m^2+\left(1-\frac{2j}{n}\right)^2}} \sin\left(\sqrt{m^2+\left(1-\frac{2j}{n}\right)^2} \, t\right) \Bigg]
	\end{align*}
	One can notice that the inhomogeneous equation is satisfied by following particular solution:
	\begin{align*}
		\widehat{u}_{n,p}(w,t) = \frac{1-\cos(w_m t)}{w_m^2}
	\end{align*}
	Therefore, the solution to the inhomogeneous IVP is given by $$\widehat{u}_{n}(w,t) = \widehat{u}_{n,h}(w,t) + \widehat{u}_{n,p}(w,t).$$
	Applying the Fourier inverse and considering we already know $u_{n,h}(x,t)$, we obtain:
	\begin{align*}
		u_n(x,t) &= u_{n,h}(x,t) + u_{n,p}(x,t) \\
				 &= \sum_{j=0}^{n} C_j(n,a) e^{ix(1-\frac{2j}{n})} \Bigg[ \cos\left(\sqrt{m^2+\left(1-\frac{2j}{n}\right)^2} \, t\right) \\
				 &+ i \frac{1-\frac{2j}{n}}{\sqrt{m^2+\left(1-\frac{2j}{n}\right)^2}} \sin\left(\sqrt{m^2+\left(1-\frac{2j}{n}\right)^2} \, t\right) \Bigg] \\
				 &+ \frac{1}{2\pi} \int_{\mathbb R} e^{ix \omega }\frac{\left(1-\cos(\sqrt{m^2+\omega^2}t)\right)}{m^2+\omega^2}d\omega.
	\end{align*}
	The limit for when $n$ goes to $\infty$ follows directly from Theorem \ref{Theorem: time evolution P1}.
\end{proof}

\begin{remark}
	The additional integral term in Theorem \ref{P2-R1} can be further evaluated using integral tables (section 3.723, equation 2. and section 3.876, equation 5. in \cite{Gradshteyn2015}),
	and becomes zero for $x>t$. For $x<t$, the integral is nonzero but does not have a simple closed-form expression. This behavior reflects causality in the Klein-Gordon equation. Specifically, for a source term at $x=\xi$ and $t=\tau$, the domain of influence is defined as:
	\begin{equation*}
		|x - \xi|< \gamma (t - \tau).
	\end{equation*}
	This is also evident from the Green's function for the 1D Klein-Gordon equation (eq. (7.4.34) in \cite{Zauderer2006PartialDE})
	\begin{equation*}
		K(x,t; \xi, \tau) = 
		\begin{cases}
			\frac{1}{2 \gamma} J_{0} \left[ \frac{c}{\gamma} \sqrt{\gamma^2 (t - \tau)^2 - (x- \xi)^2} \right],  &|x - \xi|< \gamma (t - \tau)\\
			0, &|x - \xi| > \gamma (t - \tau)
		\end{cases}
	\end{equation*}
	where in our case, $\gamma=1$ and $c=m$.
\end{remark}

\begin{theorem}\label{Theorem: Integral representation 1D Klein-Gordon}
	Given the IVP for the Klein-Gordon equation
	\begin{equation*}
		\frac{\partial^2}{\partial t^2} u(x,t) -  \frac{\partial^2}{\partial x^2}u(x,t) + m^2 u(x,t)= L(x,t), \quad \text{$-\infty < x < \infty$, $t>0$,}
	\end{equation*}
	with the initial conditions
	\begin{equation*}
		u(x,0) = f(x), \quad \partial_tu(x,0)= g(x), \quad -\infty < x < \infty.
	\end{equation*}
	The solution at any arbritary point $(x,t)$ is then given by
	\begin{align*}\label{eq: solution IVP Klein-Gordon}
		u(x,t) &= \frac{f(x-t) + f(x+t)}{2} \\
		& + \frac{1}{2} \int_{x-t}^{x+t}  J_0 \left[ m\sqrt{t^2 - (x-\xi)^2} \right] g(\xi) d\xi \\
		& - m\frac{t}{2} \int_{x-t}^{x+t} \frac{J_1 \left[ m \sqrt{t^2 - (x-\xi)^2} \right] }{ \sqrt{t^2 - (x-\xi)^2} } f(\xi) d\xi \\
		& + \frac{1}{2} \int_{0}^{t} \int_{x-(t-\tau)}^{x+(t-\tau)} L(\xi, \tau) J_0 \left[ m\sqrt{(t-\tau)^2 - (x-\xi)^2} \right] d\xi d \tau
	\end{align*}
\end{theorem}
\begin{proof}
	See \cite[eqs. (7.4.38), (7.4.39), and (7.4.41)]{Zauderer2006PartialDE}, with $\gamma = 1$ and $c = m$ in our case.
\end{proof}

\begin{theorem}\label{P5-R1}
	The time evolution of the superoscillating sequence $F_n(x,a)$ for the 1D-Klein-Gordon equation considered in Problem \ref{P3} is given by
	\begin{equation}
	\begin{aligned}
		v_n(x,t) &= \sum_{j=0}^n C_j(n,a) e^{i x \left(1-\frac{2j}{n} \right)} \left[ \cos \left( \sqrt{m^2 + \left(1-\frac{2j}{n} \right)^2} t \right) \right. \\
		& \left. + i\frac{1 - \frac{2j}{n}}{\sqrt{m^2 + \left( 1-\frac{2j}{n} \right)^2}} \sin \left( \sqrt{m^2 + \left(1-\frac{2j}{n} \right)^2} t \right)\right] +  \frac{1}{2} H(t - |x|) J_0 \left[ m\sqrt{t^2 - x^2} \right], 
	\end{aligned}
	\end{equation}
	where $H$ denotes the Heaviside function.
\end{theorem}
\begin{proof}
	Applying Theorem \ref{Theorem: Integral representation 1D Klein-Gordon}, we obtain
	\begin{align*}
		v_n(x,t) = &\sum_{j=0}^n C_j(n,a) \frac{e^{i (x-t)\left(1-\frac{2j}{n} \right)} + e^{i (x+t) \left(1-\frac{2j}{n} \right)}}{2} \\
		+ &\frac{i}{2}  \sum_{j=0}^nC_j(n,a) \left(1-\frac{2j}{n} \right)  \int_{x-t}^{x+t}  J_0 \left[ m\sqrt{t^2 - (x-\xi)^2} \right] e^{i \xi \left(1-\frac{2j}{n}\right) } d\xi \\
		- &m\frac{t}{2} \sum_{j=0}^nC_j(n,a)  \int_{x-t}^{x+t} \frac{J_1 \left[ m \sqrt{t^2 - (x-\xi)^2} \right] }{ \sqrt{t^2 - (x-\xi)^2} } e^{i \xi \left( 1-\frac{2j}{n} \right) } d\xi \\
		+ &\frac{1}{2} \int_{0}^{t} \int_{x-(t-\tau)}^{x+(t-\tau)} \delta(\xi) \delta(\tau) J_0 \left[ m\sqrt{(t-\tau)^2 - (x-\xi)^2} \right] d\xi d \tau
	\end{align*}
	The integrals in the second and third terms are known (section 6.677, equation 6. and section 6.727, equation 1. in \cite{Gradshteyn2015}, respectively). This gives
	\begin{align*}
		v(x,t) &= \sum_{j=0}^n C_j(n,a) e^{i x \left(1-\frac{2j}{n} \right)} \text{cos}\left( \left(1-\frac{2j}{n} \right) t \right) \\
		&+ i \sum_{j=0}^n C_j(n,a) e^{i x \left(1-\frac{2j}{n} \right)} \frac{1 - \frac{2j}{n}}{\sqrt{m^2 + \left( 1-\frac{2j}{n} \right)^2}} \text{sin} \left( \sqrt{m^2 + \left(1-\frac{2j}{n} \right)^2} t \right) \\
		&- \sum_{j=0}^nC_j(n,a) e^{i x \left( 1-\frac{2j}{n} \right)}  \left[ \text{cos} \left( \left( 1-\frac{2j}{n} \right) t \right) - \text{cos} \left(\sqrt{m^2+\left( 1-\frac{2j}{n} \right)^2} t \right) \right] \\
		& + \frac{1}{2} H(t - |x|) J_0 \left[ m\sqrt{t^2 - x^2} \right] 
	\end{align*}
	This simplifies to
	\begin{align*}
	v_n(x,t) &= \sum_{j=0}^n C_j(n,a) e^{i x \left(1-\frac{2j}{n} \right)} \left[ \cos \left( \sqrt{m^2 + \left(1-\frac{2j}{n} \right)^2} t \right) \right. \\
		& \left. + i\frac{1 - \frac{2j}{n}}{\sqrt{m^2 + \left( 1-\frac{2j}{n} \right)^2}} \sin \left( \sqrt{m^2 + \left(1-\frac{2j}{n} \right)^2} t \right)\right] +  \frac{1}{2} H(t - |x|) J_0 \left[ m\sqrt{t^2 - x^2} \right]. 
	\end{align*}
\end{proof}

\begin{remark}
	The solution of Theorem \ref{P5-R1}, consists of the solution to the homogeneous problem (Theorem \ref{Theorem: time evolution P1}), plus the particular solution, corresponding to the last term in Theorem \ref{Theorem: Integral representation 1D Klein-Gordon}.
	When the homogeneous solution is known, it thus suffices to add this last term to obtain the solution.
\end{remark}

\begin{theorem}\label{Theorem: time evolution problem 3.3}
The time evolution of the superoscillating sequence $F_n(x,a)$ for the 1D-Klein-Gordon equation considered in the first case of Problem \ref{NP2} is given by 
\begin{equation}
\begin{aligned}
u_n(x,t) &= \sum_{j=0}^{n} C_j(n,a) e^{ix(1-\frac{2j}{n})} \cos\left(\sqrt{m^2+\left(1-\frac{2j}{n}\right)^2} \, t\right) \\
&+ \sum_{j=0}^{n} C_j(n,b) e^{ix(1-\frac{2j}{n})} \frac{1}{\sqrt{m^2+\left(1-\frac{2j}{n}\right)^2}} \sin\left(\sqrt{m^2+\left(1-\frac{2j}{n}\right)^2} \, t\right). 
\end{aligned}
\end{equation}

If $u(x,t)$ denotes the limit case when $n$ goes to $\infty$, we obtain 

\begin{equation}\label{Equation: limitcase time evolution problem 3.3}
u(x,t)= e^{iax}\cos(\sqrt{m^2+a^2}t)+e^{ibx} \frac{1}{\sqrt{m^2+ b^2}} \sin(\sqrt{m^2+b^2}t).
\end{equation}

\end{theorem}

\begin{proof}
	After applying the Fourier transform, we obtain the same ordinary differential equation and corresponding general solution as in Lemma \ref{Lemma general solution 1D homogenous klein gordon}. The coefficients $A(w)$ and $B(w)$ can then again be determined from the initial conditions. We arrive at
	\begin{align*}
		\hat{u}(w,t) &= 2 \pi \sum_{j=0}^{n} C_j(n,a) \delta \left(w- \left(1- \frac{2j}{n} \right) \right) \text{cos}(w_m t)\\
		&+ \frac{2 \pi}{w_m} \sum_{j=0}^{n} C_j(n,b) \delta \left(w- \left(1- \frac{2j}{n} \right) \right) \text{sin}(w_m t)
	\end{align*}
	The time evolution is then found after applying the inverse Fourier transform
	\begin{align*}
		u_n(x,t) &= \sum_{j=0}^{n} C_j(n,a) e^{ix(1-\frac{2j}{n})} \cos\left(\sqrt{m^2+\left(1-\frac{2j}{n}\right)^2} \, t\right) \\
		&+ \sum_{j=0}^{n} C_j(n,b) e^{ix(1-\frac{2j}{n})} \frac{1}{\sqrt{m^2+\left(1-\frac{2j}{n}\right)^2}} \sin\left(\sqrt{m^2+\left(1-\frac{2j}{n}\right)^2} \, t\right).
	\end{align*}
\end{proof}

\begin{remark}\label{Theorem: time evolution problem 3.5}
The time evolution of the superoscillating sequence $F_n(x,a)$ for the 1D-Klein-Gordon equation considered in the second case of Problem \ref{NP2} is given by 
\begin{equation}
\begin{aligned}
u_n(x,t) &= \sum_{j=0}^{n} C_j(n,a) e^{ix(1-\frac{2j}{n})} \cos\left(\sqrt{m^2+\left(1-\frac{2j}{n}\right)^2} \, t\right) \\
&+ \sum_{j=0}^{n} C_j(n,b) e^{ix(1-\frac{2j}{n})} \frac{1}{\sqrt{m^2+\left(1-\frac{2j}{n}\right)^2}} \sin\left(\sqrt{m^2+\left(1-\frac{2j}{n}\right)^2} \, t\right). \\
&+ \frac{1}{2\pi} \int_{\mathbb R} e^{ix \omega }\frac{\left(1-\cos(\sqrt{m^2+\omega^2}t)\right)}{m^2+\omega^2}d\omega.
\end{aligned}
\end{equation}
This result follows from the same approach outlined in Theorem \ref{P2-R1}. The solution consists of the homogeneous solution from Theorem \ref{Theorem: time evolution problem 3.3} and a particular solution, which corresponds to the last additional integral term and also occurs in Theorem \ref{P2-R1}. 
\end{remark}

\subsection{An application of the Segal-Bargmann transform}

\begin{definition}
Let $m>0$ and $\omega \in \mathbb{R}$ be fixed. Define the function 
\begin{equation}
\theta_{\omega, m}(t)=\cos(\sqrt{m^2+\omega^2}t)+i\frac{\omega}{\sqrt{m^2+\omega^2}}\sin(\sqrt{m^2+\omega^2}t),\quad t\geq 0.
\end{equation}
Let $n\in \mathbb N$. Consider the $L^2$-functions defined by 
\begin{equation}
\phi_n(x,t)=\phi_{n,t}(x)=e^{-\frac{x^2}{2}}u_n(x,t), \text{ and } \phi(x,t)=\phi_t(x)=e^{-\frac{x^2}{2}}u(x,t), 
\end{equation}
where $u_n(x,t)$ and $u(x,t)$ are the solutions obtained in Theorem \ref{Theorem: time evolution P1}.
\end{definition}
\begin{remark}
Observe that 
$$\displaystyle \phi_n(x,t)=e^{-\frac{x^2}{2}}\sum_{j=0}^{n}C_j(n,a) e^{ix\left(1-2j/n\right)} \theta_{1-2j/n, m}(t),$$
and $$\phi(x,t)=e^{-\frac{x^2}{2}+iax} \theta_{a,m}(t).$$
\end{remark}

We now study the action of the Segal-Bargmann transform on the regularized solution $\phi_n(x,t)$
\begin{lemma}
The Segal-Bargmann transform $\mathcal{B}$ applied to $\phi_n(x,t)$ yields the entire function

\begin{equation}
\displaystyle (\mathcal{B}\phi_{n,t})(z)=\pi^{1/4}\sum_{j=0}^{n}C_j(n,a) \theta_{1-\frac{2j}{n},m}(t) k_{-\frac{i}{\sqrt{2}}(1-\frac{2j}{n})}(z)=:\xi_n(z,t),
\end{equation}
where $k_{-\frac{i}{\sqrt{2}}(1-\frac{2j}{n})}$ is the normalized reproducing kernel of the Fock space, explicitly given by 
$$k_{-\frac{i}{\sqrt{2}}(1-\frac{2j}{n})}(z)=e^{\frac{iz}{\sqrt{2}}(1-\frac{2j}{n})-\frac{1}{4}(1-\frac{2j}{n})^2}, \quad z\in \mathbb{C}.$$
\end{lemma}
\begin{proof}
We have
\begin{align*}
\displaystyle (\mathcal{B}\phi_{n,t})(z)&=\pi^{-1/4} \int_{\mathbb{R}} e^{-\frac{1}{2}(x^2+z^2)+\sqrt{2}zx}\phi_{n,t}(x)dx\\
&=\pi^{-1/4} \int_{\mathbb{R}} e^{-\frac{1}{2}(x^2+z^2)+\sqrt{2}zx} e^{-\frac{x^2}{2}}u_n(x,t)  dx\\
&=\pi^{-1/4} e^{-\frac{z^2}{2}}\sum_{j=0}^{n}C_j(n,a)  \theta_{1-\frac{2j}{n},m}(t)  \int_{\mathbb{R}}  e^{-x^2+\left(\sqrt{2}z+i(1-\frac{2j}{n})\right)x} dx\\\
&=\pi^{1/4} e^{-\frac{z^2}{2}}\sum_{j=0}^{n}C_j(n,a)  \theta_{1-\frac{2j}{n},m}(t)  e^{\frac{\left(\sqrt{2}z+i(1-\frac{2j}{n})\right)^2}{4}}\\
&=\pi^{1/4} \sum_{j=0}^{n}C_j(n,a)  \theta_{1-\frac{2j}{n},m}(t) e^{\frac{iz}{\sqrt{2}}(1-\frac{2j}{n})-\frac{1}{4}(1-\frac{2j}{n})^2}  \\
&=\pi^{1/4}\sum_{j=0}^{n}C_j(n,a) \theta_{1-\frac{2j}{n},m}(t) k_{-\frac{i}{\sqrt{2}}(1-\frac{2j}{n})}(z) \\
&=\xi_n(z,t).\\
\end{align*}
\end{proof}
The entire function $\xi_n(z,t)$, obtained via the Segal-Bargmann transform, yields an integral representation for the solution $u_n(x,t)$. More precisely, we have the following result:
\begin{theorem}\label{IntRep}
The solution $u_n(x,t)$ admits the following integral representation in terms of $\xi_n(z,t)$
\begin{equation}
\displaystyle u_n(x,t)=\pi^{-1/4} \int_{\mathbb C} e^{-\frac{\overline{z}^2}{2}+\sqrt{2}x\overline{z}}\xi_n(z,t)\frac{e^{-|z|^2}d\lambda(z)}{\pi}, \quad x \in \mathbb{R}, \quad t\geq 0.
\end{equation}
\end{theorem}
\begin{proof}
The Segal-Bargmann inversion formula yields
\begin{align*}
\displaystyle \phi_{n,t}(x)&=(\mathcal{B}^{-1}\xi_n)(x)\\
&=\pi^{-1/4} \int_{\mathbb C} e^{-\frac{1}{2}(x^2+\overline{z}^2)+\sqrt{2}\overline{z}x} \xi_n(z,t) \frac{e^{-|z|^2}d\lambda(z)}{\pi}. \\
\end{align*}
Since $\phi_n(x,t)=e^{-\frac{x^2}{2}}u_n(x,t)$, we obtain 

$$\displaystyle u_n(x,t)=\pi^{-1/4} \int_{\mathbb C} e^{-\frac{\overline{z}^2}{2}+\sqrt{2}x\overline{z}}\xi_n(z,t)\frac{e^{-|z|^2}d\lambda(z)}{\pi}.$$
\end{proof}
\begin{remark}
At time $t=0$, the function $\xi_n(z,t)$ simplifies to 
\begin{equation}
\xi_n(z,0)=\pi^{1/4}\sum_{j=0}^{n}C_j(n,a) e^{\frac{iz}{\sqrt{2}}(1-\frac{2j}{n})-\frac{1}{4}(1-\frac{2j}{n})^2}, \quad z\in \mathbb{C}.
\end{equation}
\end{remark}
\begin{proposition}
The superoscillating function $F_n(x,a)$  and its derivative $F'_{n}(x,a)$ admit the following integral representations

\begin{equation}
F_n(x,a)= \pi^{-1/4} \int_{\mathbb C} e^{-\frac{\overline{z}^2}{2}+\sqrt{2}x\overline{z}} \xi_n(z,0)\frac{e^{-|z|^2}d\lambda(z)}{\pi}, \quad x \in \mathbb{R},
\end{equation}

\begin{equation}
F'_{n}(x,a)= \sqrt{2}\pi^{-1/4} \int_{\mathbb C} e^{-\frac{\overline{z}^2}{2}+\sqrt{2}x\overline{z}}\frac{d}{dz}\xi_n(z,0)\frac{e^{-|z|^2}d\lambda(z)}{\pi}, \quad x \in \mathbb{R}, \quad p \geq 1.
\end{equation}
\end{proposition}
\begin{proof}
We apply Theorem \ref{IntRep} at time $t=0$, leading to $$F_n(x,a)=u_n(x,0)= \pi^{-1/4} \int_{\mathbb C} e^{-\frac{\overline{z}^2}{2}+\sqrt{2}x\overline{z}} \xi_n(z,0)\frac{e^{-|z|^2}d\lambda(z)}{\pi},$$
and \begin{align*}
F'_n(x,a)&=\frac{\partial}{\partial t}u_n(x,0)\\
&=\pi^{-1/4} \int_{\mathbb C} e^{-\frac{\overline{z}^2}{2}+\sqrt{2}x\overline{z}}\frac{\partial}{\partial t}\xi_n(z,0)\frac{e^{-|z|^2}d\lambda(z)}{\pi}.\\
\end{align*}
Using direct calculations, we observe that 

\begin{align*}
\displaystyle \frac{\partial}{\partial t}\xi_n(z,0)&=\sqrt{2}\pi^{1/4}\sum_{j=0}^{n}C_j(n,a)\frac{i}{\sqrt{2}}\left(1-\frac{2j}{n}\right) e^{\frac{iz}{\sqrt{2}}(1-\frac{2j}{n})-\frac{1}{4}(1-\frac{2j}{n})^2}\\
&=\sqrt{2} \frac{d}{d z}\xi_n(z,0).\\
\end{align*}
Finally, we conclude that $$F'_n(x,a)=\sqrt{2}\pi^{-1/4} \int_{\mathbb C} e^{-\frac{\overline{z}^2}{2}+\sqrt{2}x\overline{z}}\frac{d}{dz}\xi_n(z,0)\frac{e^{-|z|^2}d\lambda(z)}{\pi}.$$
\end{proof}

\section{Discussion}
\setcounter{equation}{0}
In this paper, we studied the evolution of superoscillations in the setting of the one-dimensional Klein-Gordon equation. We derived analytic solutions to this problem under various initial conditions and source terms. In particular, Theorem \ref{P2-R1} provides an explicit solution for the evolution of superoscillations governed by the Klein-Gordon equation in the presence of a Dirac delta source. By considering the resulting solution, we observe the appearance of an additional integral term given by

\begin{equation}
\frac{1}{2\pi} \int_{\mathbb R} e^{ix \omega }\frac{\left(1-\cos(\sqrt{m^2+\omega^2}t)\right)}{m^2+\omega^2}d\omega=\frac{1}{2\pi} \widehat{f_{t,m}}(-x),  
\end{equation}
where
\begin{equation}
f_{t,m}(\omega):=\frac{1-\cos(\sqrt{m^2+\omega^2}t)}{m^2+\omega^2}, \quad \omega \in\mathbb R, t>0.
\end{equation}
It turns out that this integral term resembles the covariance kernel function of fractional Brownian motion (fBM). To explore this connection, we introduce the following definition:
\begin{definition}
The covariance kernel of fractional Brownian motion (fBM) $B=(B_t)_{t>0}$, with Hurst parameter $H\in (0,1),$ is given by 
$$K_H(s,t)=Cov_B(s,t)=E(B_sB_t):=\frac{1}{2}(s^{2H}+t^{2H}-|s-t|^{2H}), \quad \forall s,t \in \mathbb{R}^+.$$
The case $H=\frac{1}{2}$ corresponds to the classical Brownian motion, for which the kernel covariance function becomes
$$K_{\frac{1}{2}}(s,t)=\min (s,t), \quad \forall s,t \in \mathbb{R}^+.$$
\end{definition}
\begin{remark}
The integral term in Theorem \ref{P2-R1} is closely related to the covariance kernel of fractional Brownian motion (fBM). Indeed, one can define the functions $r_{x,m}(t)$ as

$$r_{m,x}(t):=\frac{1}{2\pi} \int_{\mathbb R} e^{ix \omega }\frac{\left(1-\cos(\sqrt{m^2+\omega^2}t)\right)}{m^2+\omega^2}d\omega=\frac{1}{\pi}  \int_{\mathbb R} e^{ix \omega }\frac{\sin^2(\frac{t}{2}\sqrt{m^2+\omega^2})}{m^2+\omega^2}d\omega,\quad t\in \mathbb{R}^+.$$

In the special case where $x=m=0$, the function $r_{0,0}(t)$ simplifies to
$$r_{0,0}(t)=\frac{1}{2\pi} \int_{\mathbb R} \frac{1-\cos(\omega t)}{\omega^2}d\omega=\frac{r(t)}{2},$$
where $r(t)=|t|$, which is a key function associated with the classical Brownian motion ($H=\frac{1}{2}$). 
\end{remark}
Inspired by \cite{AAL2010}, we introduce the following: 
\begin{definition} 
We define an extension of the Brownian motion using the covariance kernel function:
\begin{equation}
\mathsf{K}_{m,x}(s,t)=Cov_X(s,t)=E(X_sX_t):=r_{m,x}(s)+\overline{r_{m,x}(t)}-r_{m,x}(s-t) \quad \forall s,t \in \mathbb{R}^+.
\end{equation}
\end{definition}
\begin{proposition}
When $m=x=0$, the kernel function $\mathsf{K}_{m,x}(s,t)$ reduces to the classical Brownian motion $$K_{\frac{1}{2}}(s,t)=\min (s,t).$$
\end{proposition}
\begin{proof}
Let $s,t \in \mathbb{R}^+$. We compute
\begin{align*}
\displaystyle \mathsf{K}_{0,0}(s,t)&=r_{0,0}(s)+r_{0,0}(t)-r_{0,0}(s-t)\\ 
&=\frac{1}{2}\left(r(s)+r(t)-r(s-t)\right)\\
&=\frac{1}{2}\left(|s|+|t|-|s-t| \right)\\
&=\min (s,t)\\
&=K_{\frac{1}{2}}(s,t). \\
\end{align*}
\end{proof}
\begin{remark}
An interesting particular case is the following 
$$\mathsf{K}_{m,0}(s,t)= r_{m,0}(s)+r_{m,0}(t)-r_{m,0}(s-t),$$
where 
$$\displaystyle r_{m,0}(t)=\frac{1}{2\pi} \int_{\mathbb R} \frac{1-\cos(\sqrt{m^2+\omega^2}t)}{m^2+\omega^2}d\omega=\frac{1}{\pi}  \int_{\mathbb R} \frac{\sin^2(\frac{t}{2}\sqrt{m^2+\omega^2})}{m^2+\omega^2}d\omega, \quad \forall t\in \mathbb{R}^+.$$
The integral on the right hand side is likely related to the family of integrals considered by von Neumann and Schoenberg in \cite[Theorem 1, p. 229]{VNS1941}.
\end{remark}

\section*{Acknowledgments}
 
The research of Kamal Diki is supported by the Research Foundation–Flanders (FWO) under grant number 1268123N.

\section*{Data avaliability}
There is no data to support the findings of this study

\end{document}